\documentclass[twocolumn]{aastex63}

\usepackage{hyperref}
\usepackage{verbatim}
\graphicspath{{./}{figures/}}
\received{Dec. 11, 2020}
\revised{Jan. 11, 2021}
\accepted{Jan. 19, 2021}
\submitjournal{ApJL}
\usepackage{amsmath}
\usepackage{amsfonts}
\shorttitle{PSR J1810+1744}
\shortauthors{Romani et al.}
\defcitealias{romani2011orbit}{RS11} 	
\defcitealias{Kandel_2020}{KR20}
\defcitealias{2017ApJ...845...42S}{SR17}
\begin{document}

\title{PSR J1810+1744: Companion Darkening and a Precise High Neutron Star Mass}

\correspondingauthor{R.W. Romani}
\email{rwr@astro.stanford.edu}

\author[0000-0001-6711-3286]{Roger W. Romani}
\affil{Department  of  Physics,  Stanford  University,  Stanford,  CA 94305, USA}
\author[0000-0002-5402-3107]{D. Kandel}
\affil{Department  of  Physics,  Stanford  University,  Stanford,  CA 94305, USA}
\author[0000-0003-3460-0103]{Alexei V. Filippenko}
\affil{Department of Astronomy, University of California, Berkeley, CA 94720-3411, USA}
\affil{Miller Institute for Basic Research in Science, University of California, Berkeley, CA 94720, USA}
\author{Thomas G. Brink}
\affil{Department of Astronomy, University of California, Berkeley, CA 94720-3411, USA}
\author{WeiKang Zheng}
\affil{Department of Astronomy, University of California, Berkeley, CA 94720-3411, USA}

\begin{abstract}
Keck-telescope spectrophotometry of the companion of PSR J1810+1744 shows a flat, but asymmetric light-curve maximum and a deep, narrow minimum. The maximum indicates strong gravity darkening near the $L_1$ point, along with a heated pole and surface winds. The minimum indicates a low underlying temperature and substantial limb darkening. The gravity darkening is a consequence of extreme pulsar heating and the near-filling of the Roche lobe. Light-curve modeling gives a binary inclination $i = 65.7^\circ \pm 0.4^\circ$. With the Keck-measured radial-velocity amplitude $K_{\rm c} = 462.3\pm 2.2\,$ $\mathrm{km\,s}^{-1}$, this gives an accurate neutron star mass $M_{\rm NS} = 2.13\pm0.04\,M_\odot$, with important implications for the dense-matter equation of state. A classic direct-heating model, ignoring the $L_1$ gravitational darkening, would predict an unphysical $M_{\rm NS} > 3\, M_\odot$. A few other ``spider" pulsar binaries have similar large heating and fill factor; thus, they should be checked for such effects.
\end{abstract}

\keywords{pulsars:  general — pulsars: individual (PSR J1810+1744)}

\section{Introduction} \label{sec:intro}

PSR J1810+1744 (hereafter J1810) was discovered by \citet{2011AIPC.1357...40H} as a $P_s=1.7$\,ms, ${\dot E}=4 \times 10^{34}\,I_{45}\, {\rm erg\,s^{-1}}$ pulsar in a 350\,MHz Green Bank Telescope (GBT) search of bright unidentified {\it Fermi} $\gamma$-ray sources. The dispersion measure $DM=39.7\,{\rm cm^{-3}\,pc}$ indicates a distance of 2.4\,kpc \citep{yao2017new} or 2.0\,kpc \citep{2002astro.ph..7156C}. It has a double-peaked $\gamma$-ray pulse with flux of $2.3\times 10^{-11}\,{\rm  erg\,cm^{-2}\, s^{-1}}$ \citep{2013ApJS..208...17A}. J1810 is in a $P_b=3.6$\,hr binary, with a relatively large $x_1=a\, {\rm sin}\, i = 0.0953$\,lt-s giving a companion mass function $4.2 \times 10^{-5}\,M_\odot$. Thus, for neutron star (NS) mass $1.5\,M_\odot$, this gives $m_c \approx 0.05\,M_\odot/{\rm sin}\,i$, relatively heavy for a black-widow (BW) companion.

The companion was first identified optically by \citet{breton2013discovery} via 2010 Gemini GMOS-N $g$-band and $i$-band imaging. Although the light curves were sparse, a fit to the data estimated an inclination of $i\approx 48^\circ \pm 7^\circ$. \citet{schroeder2014observations} presented more extensive $BVR$ photometry of the bright phases of J1810 using the MDM Observatory. They find some disagreement with the $g$ photometry of \citet{breton2013discovery}, and quote an orbital inclination of $i\approx 56^\circ \pm 3^\circ$ and a small phase shift of the optical maximum from pulsar inferior conjunction.

\begin{figure*}[t!!]
    \centering
    \vspace*{-5mm}\hspace*{-5mm}\includegraphics[scale=0.95]{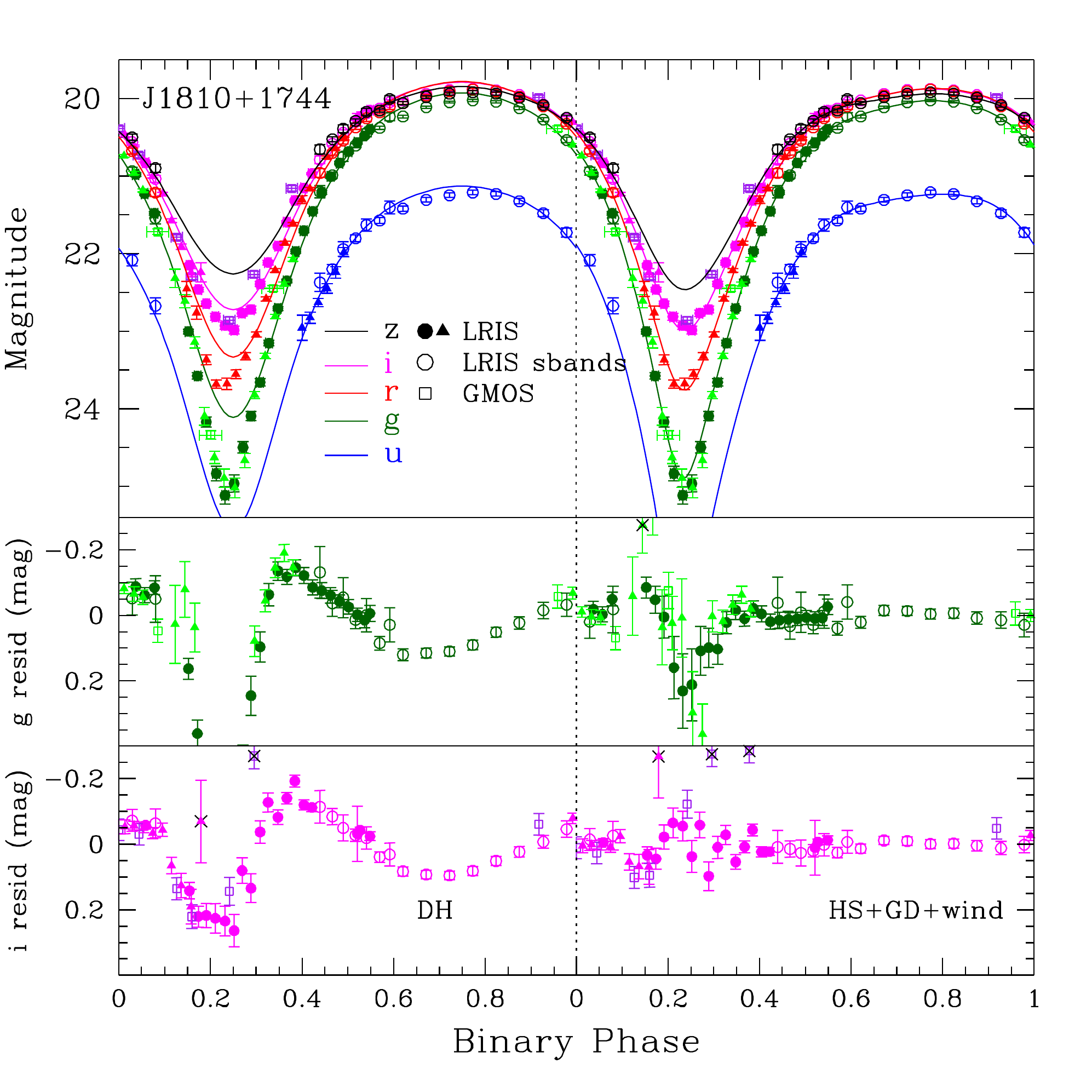}
    \vspace*{-0mm}
    \caption{Calibrated {\it ugriz} light curves from J1810 Keck LRIS spectroscopy (2017), Keck photometry (2020), and GMOS photometry (2010). Two cycles are shown and these data are available online as DbF. The model in the first cycle is the best-fit direct heating model; the second cycle shows the best-fit HS+GD+wind model. Residuals to these two models are shown for $g$ and $i$ in the lower two panels. A few outlier points (marked with black x symbols) are excluded from the fit (see text). See Table \ref{tab:lc_fit} for the model fit parameters.} 
    \label{fig:LC}
\end{figure*}

Here we describe  Keck optical spectroscopy and photometry of J1810 that allows improved modeling of the companion heating pattern. This is important since, in addition to direct pulsar $\gamma$-rays, the companion heating can be affected by photons from the system's intrabinary shock \citep[IBS;][]{2016ApJ...828....7R}, by IBS particles ducting along magnetic field lines to companion poles \citep{2017ApJ...845...42S}, and by heat transfer from a global wind \citep[][hereafter \citetalias{Kandel_2020}]{Kandel_2020} or general surface diffusion \citep{voisin2020model}. These effects distort the light curve, affecting the measurement of the system viewing angle and shifting the companion center of light (CoL) and radial-velocity center from the center of mass (CoM), and can vary from epoch to epoch \citep{2020ApJ...903...39K}, affecting our estimate of the NS mass.

\section{Observations}

We collected  J1810 spectra with the Keck-I 10\,m telescope and LRIS \citep{oke1995keck} for $23\times 600$\,s on UT August 26, 2017 (MJD 57991.285--57991.452), covering 1.1 binary orbits. We used the 5600\,\AA\, dichroic, the 600\,l/4000\,\AA\, blue grism, and the 400\,l/8500\,\AA\, red grating, covering $\sim 3300$--10,500\,\AA\ with dispersions of 0.63\,\AA\,pixel$^{-1}$ (blue side) and 1.2\,\AA\,pixel$^{-1}$ (red side). The atmospheric dispersion corrector (ADC) allowed us to rotate the $1^{\prime\prime}$-wide slit away from the parallactic angle \citep{Fil82} to simultaneously monitor a nearby brighter F8 star with known Pan-STARRS2 (PS2) magnitudes, to monitor the system throughput and wavelength solution between frames. In particular, since this PS2 monitor star has known and stable magnitudes, we integrate the spectra over the SDSS standard $ugriz$ filter bands using the {\tt sbands} {\tt IRAF} script, and calibrate with the catalog magnitudes (converted to the SDSS system using the prescription of \citealt{finkbeiner2016hypercalibration}) to obtain light curves with absolute fluxes, up to a possible small grey shift from differential slit losses. 

This procedure worked well; the  {\tt sbands} magnitudes near maximum brightness are stable and, despite substantially decreased throughput toward the end of the observation, the fluxes match well in the overlap phases ($\phi_B=0.45$--0.6). Photometric errors have been estimated by scaling to the raw spectrum signal. Toward minimum brightness, however, the errors are large and the cadence too slow for a light curve of suitable quality. 

Thus, we also used LRIS for direct dual-band photometry, collecting a sequence of $g/I$ images on UT August 21, 2020 (MJD 59083.38--59083.46) as well as $g/R$, $U/R$, and $g/I$ images on UT Sep. 17, 2020 (MJD 59100.21--59100.38). These had variable 0.8--1.1$^{\prime\prime}$ seeing. We performed forced photometry of the companion and a grid of nearby PS2 catalog stars in $1.35^{\prime\prime}$ diameter apertures. The catalog magnitudes, converted to the SDSS scale, were used to calibrate the companion light curve and estimate systematic errors. In the $0.4^{\prime\prime}$ full width at half-maximum intensity (FWHM) GMOS-N minimum images of \citet{breton2013discovery}, a faint star lies $0.83^{\prime\prime}$ from the pulsar. We measured fluxes of this star and the pulsar companion in all the GMOS-N $g/i$ frames, using comparison stars from the PS2 catalog, and subtract the estimated contaminator flux in the LRIS photometry apertures. We account for the LRIS/GMOS filter and monitor star differences with small shifts ($-0.2$\,mag in $g$, $-0.15$\,mag in $i$), and find that the points agree well with our LRIS-derived fluxes except in the phase range $0.3<\phi_B < 0.4$, where the GMOS fluxes are $\sim 0.3$\,mag brighter than in the LRIS data. This is apparently the phase range of a heated spot (see below), and the heating may have been stronger in 2010. We also bring the LRIS spectrophotometry to the imaging photometry scale with a gray shift of $\sim +0.2$\,mag, suggesting that the monitor-star slit losses were $\sim 20$\% larger than those for the companion.

For the light-curve fitting, we used points with total errors (photometric and systematic calibration uncertainties, added in quadrature) $\sigma_m < 0.15$\,mag. However, we dropped two $g$ points in twilight and one $i$ point taken as the source was setting, and excluded the {\tt sbands} photometry at $0.1<\phi_B<0.4$ (low signal-to-noise ratio and large $\Delta \phi$). Altogether we have 23 $u$, 60 $g$, 34 $r$, 51 $i$, and 17 $z$ measurements (Fig.\,\ref{fig:LC}). Both our new light curve and the MDM $BVR$ photometry of \citet{schroeder2014observations} show a rather broad, flat maximum, with a slight brightening to later phases and little color variation. Our Keck imaging with its fine cadence shows a narrow light-curve minimum, reaching $g>25$\,mag. The night-phase light curve is significantly asymmetric, with the minimum at $\phi_B<0.25$.

The extinction in this direction is estimated from the three-dimensional dust maps of \citet{2018MNRAS.478..651G}, reaching its maximal $E(g-r)=0.12\pm 0.02$\,mag ($A_V =0.39 \pm 0.07$\,mag) by 1.1\,kpc.

Balmer absorption lines dominate the companion spectrum through the bright half of the orbit (spectral class $\sim$A2 at maximum brightness). Accordingly, we initially measure the radial-velocity (RV) amplitude by cross-correlation with an A2 template. The correlation during the bright ``day" phases gives RV uncertainties as small as 5\,km\,s$^{-1}$. Although the effective temperature drops dramatically, the correlation persists with plausible velocities (with large uncertainty) for almost all night-phase spectra. We have chosen to focus on the measurements with the best correlation coefficient ($R>10$). This gives 15 RV measurements covering phase $\phi_B=0.45$--1.05. We have removed the nominal systemic velocity. A simple sinusoidal fit to these data gives $K_{\rm obs} = 426.9 \pm 3.4\,{\rm km \, s^{-1}}$ with $\chi^2/$DoF = 2.3 (Fig. 2).

\begin{figure}[t!]
\centering
\hspace*{-5mm}\includegraphics[scale=0.45]{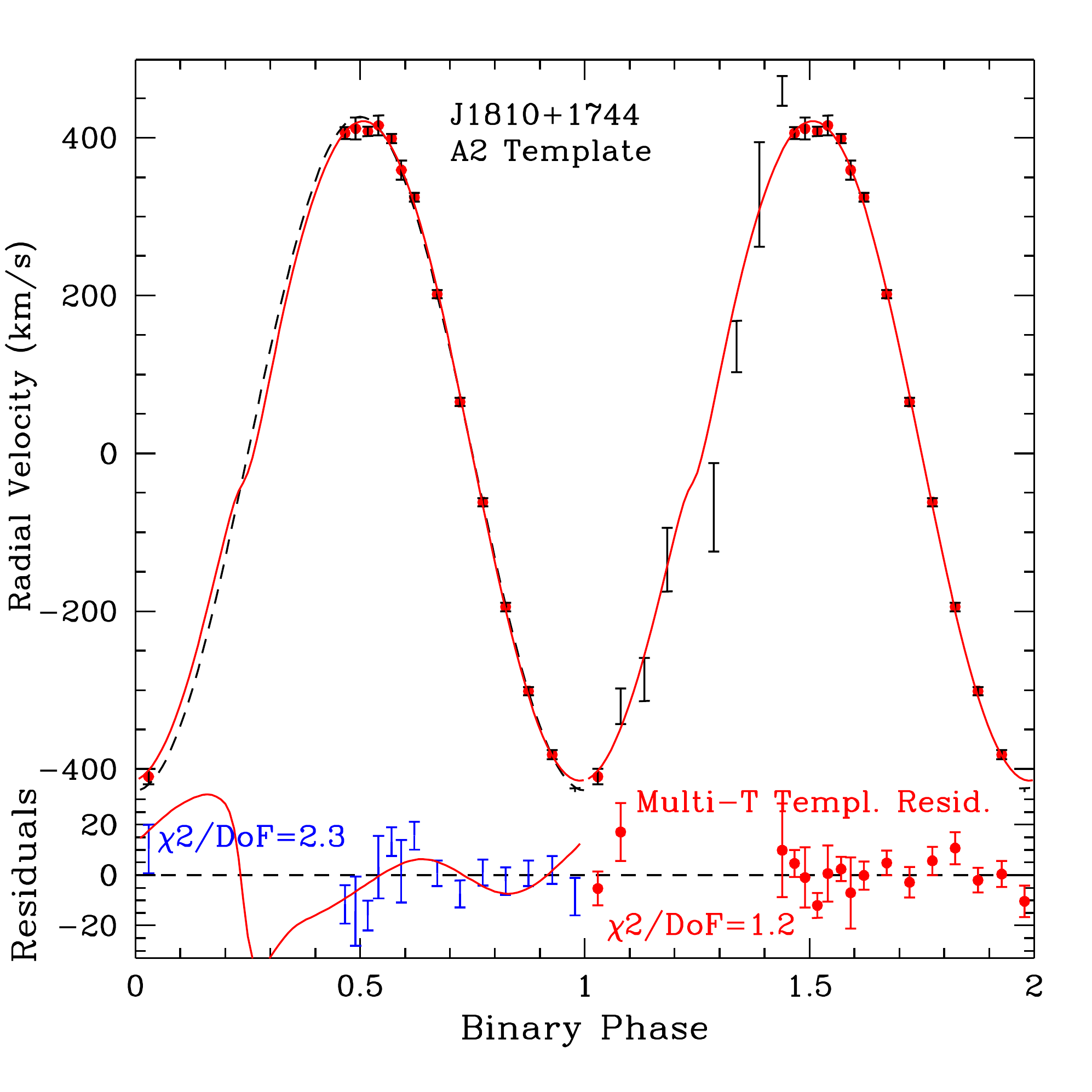}
\caption{Radial-velocity (RV) measurements for J1810. The first cycle shows 15 high-correlation ($R>10$) measurements against an A2 template (velocities available on-line as DbF). These are fit to a simple sinusoid (dashed curve) with the fit residuals indicated at bottom. The curve at bottom shows the sinusoid residuals expected for the best HS+GD+wind photometric fit; the fit residuals follow this curve. The second cycle includes low-significance velocity correlations from the night half of the orbit. The solid line is again the best-fit photometric model. For this model, the phase spectral templates provide 17 measurements with $R>10$, decreased residuals (shown in the bottom section), and an improved $\chi^2$/DoF = 1.2 fit.}
\label{fig:RV}
\end{figure}

\section{Photometric/Radial-Velocity Fitting}

Our fits are performed with an outgrowth of the {\tt ICARUS} light-curve model \citep{breton2012koi} with additions described by KR20 and \citet{2020ApJ...903...39K}. An additional update replaced the simplified limb-darkening laws in the base code with the more detailed limb-darkening coefficients computed  by \citet{2011A&A...529A..75C} for two models, the ATLAS and PHOENIX atmospheres. We generally find that the ATLAS coefficients perform better. Note that gravity and limb darkening serve to rescale the local fluxes; we take care to integrate the emergent flux to determine the total nonthermal pulsar heating $L_{\rm H}$ and the thermal base emission (characterized by the night-side temperature $T_{\rm N}$).

All fits find that the companion is very close to Roche-lobe filling, so we set the fill factor at $f_1=0.99$. We also find that, when left free, the extinction prefers values larger than estimated from the dust maps. We have chosen to fix this at $A_V=0.5$\,mag for our basic fits, about $1.5\sigma$ larger than the dust estimate, and note below when a free fit gives a significantly different value. A standard direct heating (DH) fit gives $i\approx 53^\circ$; similar results were obtained by \citet{schroeder2014observations}. The cause is the model's attempt to produce a flat maximum  via low inclination, $i$. With our newly measured $K$, this value of $i$ implies an unphysical $M_{\rm N} > 3\,M_\odot$ (Table \ref{table:fit}).

Asymmetry of the light-curve maximum and bluer colors leading the peak indicate that there is extra heat at this phase. We model this with a Gaussian hot spot (HS) of radius $r_{\rm hs}$ centered at $(\theta_{\rm hs}, \phi_{\rm hs})$, heated to $(1+A_{\rm hs})$ times the local star temperature. In the picture of \citet{2017ApJ...845...42S}, such hot spots are caused by relativistic particles ducting from the pulsar wind and IBS to companion magnetic poles. Note that the excluded GMOS data points near the HS phase ($\phi_B \approx 0.4$)  indicate that this phase was brighter and bluer in 2010; we speculate that the particle flux ducting to the HS varies between epochs (as seen for PSR J2339$-$0533 by \citealt{2020ApJ...903...39K}), and was stronger at the GMOS epoch. 

We next amend the gravity darkening \citep[GD;][]{2012A&A...547A..32E}. While the standard ICARUS code applies GD to the underlying companion, it does not apply to the heated companion face. J1810 has a Roche lobe fill factor $f_1 \approx 1$ (giving very low $g$ near the nose) and temperatures on the heated face exceed $10,000$\,K, so GD effects can be substantial, We use a simple prescription $T' = T (g/g_0)^\beta$. For the effectively radiative photosphere of the strongly heated day face we use $\beta=0.25$ (as usual $\beta\approx 0.08$ applies for the low-$T$ convective atmosphere on the back side), with the scaling $g_0$ taken from the dawn equator point. The pre-GD temperature distribution includes such effects as hot spots and surface heat redistribution, if any.

To produce the broadening and slight gradient across the maximum, some flux must be moved to the trailing side. We find that a simple global wind, with heat advected along latitude lines (see KR20 for model details), greatly helps in matching the maximum shape. An alternative is to invoke heat diffusion away from the companion nose, as described by \citet{voisin2020model}. This does broaden the peak, but without the asymmetry reproduced with the HS+GD+wind model. Table \ref{table:fit} shows the nested light curve fits as these effects are turned on; the large $\chi^2$ decreases indicate that the added parameters are highly significant.

With these effects included, we have a much better companion light-curve fit; wind and GD flatten the day-face temperature, while a hot spot reproduces the observed peak asymmetry. The very narrow deep minimum is not fully captured. The model minimum is sensitive to the limb-darkening coefficients: use of the PHOENIX model-derived coefficients produces significantly worse fits than those with the ATLAS model coefficients, possibly due to a different treatment at the low temperatures found on the `night' side. Although not employed here, a further small reduction in $\chi^2$ can be found by decreasing the limb darkening at low temperature. Thus, additional physical ingredients are needed and might be probed with more data and improved atmosphere modeling.

\begin{deluxetable*}{lcccc|ccc}
\tabletypesize{\footnotesize}
\tablewidth{0pt}
\tablecaption{Light Curve and Template Spectral Fit Results for J1810\label{tab:lc_fit}}
\tablehead{
\colhead{Parameters} & \colhead{DH} & \colhead{HS} & \colhead{HS+GD} & \colhead{HS+GD+wind} & \colhead{HS+GD/$A_V$\tablenotemark{*}} & \colhead{HS+GD+diff} & \colhead{2HS+GD}}
\startdata
$i\, (\mathrm{ deg})$ & $52.9^{+0.8}_{-0.7}$  & $56.6^{+0.7}_{-0.7}$  & $61.7^{+0.7}_{-0.6}$  & $65.7^{+0.4}_{-0.4}$ &$66.1^{+0.5}_{-0.5}$ &$68.0^{+1.0}_{-0.9}$ &$62.7^{+0.4}_{-0.4}$ \\
$L_{\mathrm{H}}\,/10^{34}\,(\mathrm{erg/s})$ & $7.04^{+0.23}_{-0.23}$  &$6.31^{+0.17}_{-0.17}$  & $6.19^{+0.13}_{-0.13}$ & $6.00^{+0.06}_{-0.06}$ & $6.75^{+0.25}_{-0.24}$ &$10.9^{+0.93}_{-0.90}$  &  $5.91^{+0.07}_{-0.07}$\\
$T_{\rm N}$\,(K) & $3320^{+100}_{-100}$ & $3440^{+60}_{-70}$ &$3550^{+50}_{-50}$ & $3470\pm 25$& $3610^{+35}_{-40}$ &$2430^{+190}_{-175}$ &$3560^{+25}_{-25}$  \\
$d_{\rm kpc}$ & $3.21\pm 0.07$  & $3.10\pm 0.06$  & $3.02\pm 0.03$  & $3.03\pm 0.01$ & $2.91\pm 0.02$ & $3.06\pm 0.05$ &$3.06\pm 0.02$\\
 $\theta_{\rm hs}\, (\mathrm{deg})$ & - &$108.1^{+2.2}_{-2.3}$ & $112.6^{+2.1}_{-2.1}$ & $100.6^{+1.1}_{-1.3}$ & $118.4^{+2.0}_{-2.1}$  & $117.5^{+2.1}_{-2.2}$& $106.6^{+0.9}_{-0.9}$\\
$\phi_{\rm hs}\, (\mathrm{deg})$  & - & $43.3^{+7.6}_{-6.9}$ & $46.0^{+5.8}_{-6.1}$ & $22.7^{+4.2}_{-4.8}$ & $51.4^{+4.0}_{-4.2}$  &$53.8^{+3.6}_{-3.9}$ & $30.6^{+4.0}_{-3.8}$\\
$\mathcal{A}_{\rm hs}$  & - & $1.9^{+0.4}_{-0.3}$ & $1.5^{+0.3}_{-0.2}$& $1.2\pm 0.1$ & $1.3\pm 0.1$ & $2.1\pm 0.4$& $1.2\pm 0.1$ \\
$\mathcal{A}_{\rm hs2}$ & - & - & - &-&-& -& $0.8\pm 0.1$ \\
 $r_{\rm hs}\, (\mathrm{deg})$ & - & $19.7^{+3.4}_{-3.1}$ &$18.8^{+2.6}_{-2.6}$& $25.7^{+1.2}_{-1.1}$ & $18.5^{+2.0}_{-1.8}$& $16.6^{+1.7}_{-1.7}$ & $20.2^{+1.5}_{-1.3}$ \\
 $\epsilon$ & - & - & - & $-0.104\pm 0.005$ & - & -&-\\
 $\sigma_{\rm diff}$ & - & - & - & - & - &$8.9^{+1.8}_{-1.6}$ & -\\
 $\chi^2/{\rm DoF}$ & 3465/179 &1750/175 & 886/175 &248/174 & 535/173  & 523/174& 270/174\\
 \hline
 \hline
 $K_{\rm CoM}$(km/s) & $474.9\pm 2.0$ & $473.0\pm 1.9$ & $468.3\pm 2.0 $ & $462.3\pm 2.2$ & $462.5\pm 2.6$ & $463.2\pm 2.1$ &  $465.0\pm 2.1$ \\
 $M_{\rm NS}\,(M_\odot)$ & $3.42\pm 0.09$ & $2.95\pm 0.07$ & $2.45\pm 0.05$ & $2.13\pm 0.04$& $2.11\pm 0.05$ & $2.03\pm 0.08$  & $2.33\pm 0.04$ \\
 $M_{\rm C}\,(M_\odot)$ &  $0.101\pm 0.002$ & $0.087\pm0.002$ & $0.073\pm0.001$ & $0.065\pm 0.001$& $0.064\pm 0.001$ & $0.062\pm 0.003$  & $0.070\pm 0.001$\\
 $\chi^2$ &16.53 &16.08 & 16.69 &18.28 &19.41 & 18.38 &16.52\\
 $\chi^2/{\rm 15\,DoF}$ &1.10& 1.07 & 1.11 & 1.22  &1.29 & 1.23 &1.10\\
 \enddata
 \tablenotetext{*}{Fit $\beta=0.46\pm0.02$, $A_V=0.60\pm0.03$\,mag}
 \label{table:fit}
\end{deluxetable*}

We next compare the Keck spectroscopy with the photometric-fit models. As emphasized by \citet{linares2018peering} and discussed by \citetalias{Kandel_2020}, different species' temperature sensitivities cause varying line equivalent widths (EWs) across the face of the companion. As the model parameters change, the surface heating changes and the line EWs vary. Since our spectra are strongly Balmer dominated, except at $\phi_B\approx 0.25$, a simple temperature-dependant weight for the Balmer EW allows a fit to the A2 cross-correlation measurements of the RV parameters, while allowing the heating model to vary. We do not fit RVs simultaneously with the photometric data; however, the RV fits for the CoM velocity and the corresponding masses are marginalized over the parameters of every tenth model from the last 1000 of the photometric nested sampling chain \citep{2014A&A...564A.125B}, sampling $2\sigma$ uncertainties. Thus, we obtain errors on the CoM $K$ ($\sigma_K$) and total uncertainties on the component masses, including the uncertainties in the photometrically determined parameters (e.g., inclination $i$). The photometric parameter contribution $\sigma_{\rm phot}$ can be isolated by propagating the contribution of $\sigma_K$ and subtracting in quadrature: $\sigma_{\rm phot}/M_{\rm NS} = [(\sigma_{M_{\rm NS}}/M_{\rm NS})^2 - (3\sigma_K/K)^2]^{1/2}$. 

Of course, the companion presents multi-temperature spectra, varying with phase, so the initial A2 templates are imperfect representations. However, once a photometric model has fit parameters, we can use this model to compute in detail, for the model CoM $K$, the expected spectrum collected during each observational integration, including all surface temperature and log\,$g$ effects and the Doppler shifts associated with each region of the surface. These integrated spectral models, computed with $\sim 10$ times the data spectral resolution, should better represent the line strengths and Doppler distortions of the companion spectra than the original A2 templates. We can thus cross-correlate the observed spectra with these multi-temperature templates to obtain corrections to the model's CoM RV at each phase. Indeed, with these templates 17 spectra now have correlation $R>10$. The cross-correlation residuals to the model RVs can be fit with a simple sinusoid to give improved measurements of the CoM RV $K$, with smaller $\chi^2$ and $\sigma_K$ uncertainties. Adding in quadrature the $\sigma_{\rm phot}$ contributions measured with the A2 template fits, we obtain the final mass error estimates. These are reported in the bottom section of Table~1, and the velocity residuals for spectral template fitting of the best-fit photometric model are shown in the lower panel of Figure \ref{fig:RV} (second cycle).

Note that as the heating model varies, the spectral template and hence cross-correlation RV for each observation will change. Recomputing the model spectra and remeasuring the velocities at each fit step would be computationally prohibitive. Thus, our hybrid scheme, fitting for a fixed A2 template to capture the RV correlations with the (photometrically determined) heating parameters, while using the composite template model spectra for the final CoM RV amplitude $K_{\rm CoM}$, allows us to capture both heating model systematics and an accurate spectral representation for the final velocities. The RV changes from the A2-determined estimates are only a few km\,s$^{-1}$, so this partition of the fits should be robust.

As noted, with the DH model assumed in most other spider binary studies, the large observed velocity amplitude of J1810 gives an unphysically high NS mass. Successively including companion HS, GD, and wind effects moves the CoL back toward the CoM, and (as seen above) increases the best-fit inclination. This substantially decreases the inferred NS mass. The $\chi^2$/DoF of the RV fit with the HS+GD+wind fit is also lower than that of a simple sinusoidal RV fit (Fig. 2), implying that the heating distortions of the model are reflected in both the photometric and spectroscopic data. However, the $\chi^2$ differences between the various heating models are not significant in the RV fits; all models are acceptable and the HS+GD+wind model has an RV-fit probability only 27\% lower than that of the DH model. Model selection is thus based on the photometric fit. 

\section{Model-Fit Comparisons}

We have demonstrated that a simple direct (photon) heating model does not adequately describe the light curve of J1810. However, with the effects of a localized hot spot, global winds, and an improved treatment of gravity and limb darkening, we get a reasonable photometric fit with $\chi^2/{\rm DoF}=1.4$ dominated by local unmodeled effects at minimum brightness. We have attempted to see if other models can do as well, but the (HS+GD+wind) model remains the best fit. Before describing these models, it is interesting to note the common features of all viable fits. First, the companion is very close to filling its Roche lobe. Next, all successful models require that the companion ``nose" near L1 be appreciably gravity darkened.

All successful models also require extra heating in a spot past the dawn terminator, giving excess flux at phase $\phi_B\approx 0.4$. Such features suggest that the companion supports a large global field to channel particles to magnetic poles. Such spots are commonly found on the redback systems with larger ($\ge 0.08\,M_\odot$), core-fusion-supporting companion masses, but are not usually prominent in true BWs. It is interesting that J1810's companion is the most massive known among the BWs (although apparently below the core-fusion threshold). We have weak evidence that the heating flux varies, with larger values during the 2010 GMOS observations. It will be interesting to determine whether, as for some redbacks \citep[see][]{2020ApJ...903...39K}, the heated spots vary in both flux and position. Precision multicolor light curves following such variation could further refine the underlying heating pattern. 

Another feature common to all the fits is a large radiation (DH) flux, $L_{\rm H} \ge 6 \times 10^{34}$\,erg\,s$^{-1}$. At first sight, this might seem in conflict with the observed {\it Fermi} flux, which for an isotropic emitter at our fit distance implies $L_\gamma =2.5 \times 10^{34}$\,erg\,s$^{-1}$. However, modern high-altitude $\gamma$-ray emission models, from the outer magnetosphere or near wind-zone, direct the radiation toward the spin equator (and companion) more than the isotropic $L_{\rm H}$ estimate in Table \ref{tab:lc_fit}. Thus, the true $L_{\rm H}$ is lower by a model-dependant beaming factor \citep[see][]{draghis2019multiband}. Also, one might worry that $L_{\rm H}$ exceeds the spin-down power ${\dot E}=4 \times 10^{34}\, I_{45}\, {\rm erg\,s^{-1}}$. However, the beaming correction helps, and one should remember that with the large mass inferred here we expect $I_{45} \approx 2$--2.5 for the stiff equations of state that allow $2\,M_\odot$ NSs. 

Table \ref{tab:lc_fit} shows the dramatic fit improvements as HS, GD, and wind effects are added, justifying our relatively complex model. However, one can imagine other physical effects producing a similar heat distribution; we show three such models in the second section of the Table, with similar parameter count to our best-fit (HS+GD+wind) model. All require an HS at nearly the same location. For example, if we free gravity darkening and extinction values (HS+GD/$A_V^*$), $A_V$ increases further above the dust estimate, the fit $\beta=0.46$ nears the largest plausible values \citep{2011A&A...529A..75C}, and yet $\chi^2$ is still two times that of our best-fit model. A similar $\chi^2$ can be achieved by replacing the global wind by heat diffusion away from local maxima (HS+GD+diff). While this diffusion broadens the peak, it does not duplicate the peak gradient. Notice that other parameters (e.g., inclination $i$, HS location) are very similar for these models. With a similar heating pattern, it is unsurprising that best-fit pulsar masses are within 1$\sigma$. 

The last alternate model posits that the companion magnetic field is dipolar, with an antipodal HS in the southern hemisphere, which serves to add flux, broadening the light-curve maximum. This spot is actually on the companion day side, below the equator, and the best-fit model drives to slightly lower inclination $i$ to decrease its contribution to the light curve. Although this has the best $\chi^2$ of the alternative models with similar parameter count, we do not favor it; the second (day-side) spot actually has a lower fit flux (we expect the particle ducting there to be stronger) and the suppressed $i$ increases the fit mass by $3.4\sigma$.

\section{Mass Implications and Conclusions}

Including the improved RV measurements with composite template spectra, the statistical error on the NS mass of our best-fit model, including all spectroscopic and light-curve effects, is quite small at $0.04\,M_\odot$. For related single HS physical models, the mass changes are small ($\sim 1 \sigma$) but the $\chi^2$ values are substantially worse. The 2HS model gives a higher though possibly acceptable $\chi^2$, but requires a $0.2\,M_\odot$ mass increase. We conclude that J1810 is a heavy NS at $2.13\pm0.04\,M_\odot$, with a robust lower limit; poorer, but possibly acceptable, models allow masses as large as $2.3\,M_\odot$.

As noted above, J1810 has the largest companion mass among the true BWs. Whether this represents an early stage of companion evaporation or an unusual early termination of the mass-transfer phase is unclear. The high heating luminosity and large fill factor raise the possibility that the pulsar may transition back to an accretion phase. If so, it must be on a decades timescale as J1810 does not show strong variability in the {\it Fermi} flux record.

\begin{figure}[t!]
\centering
\hspace*{-5mm}\includegraphics[scale=0.45]{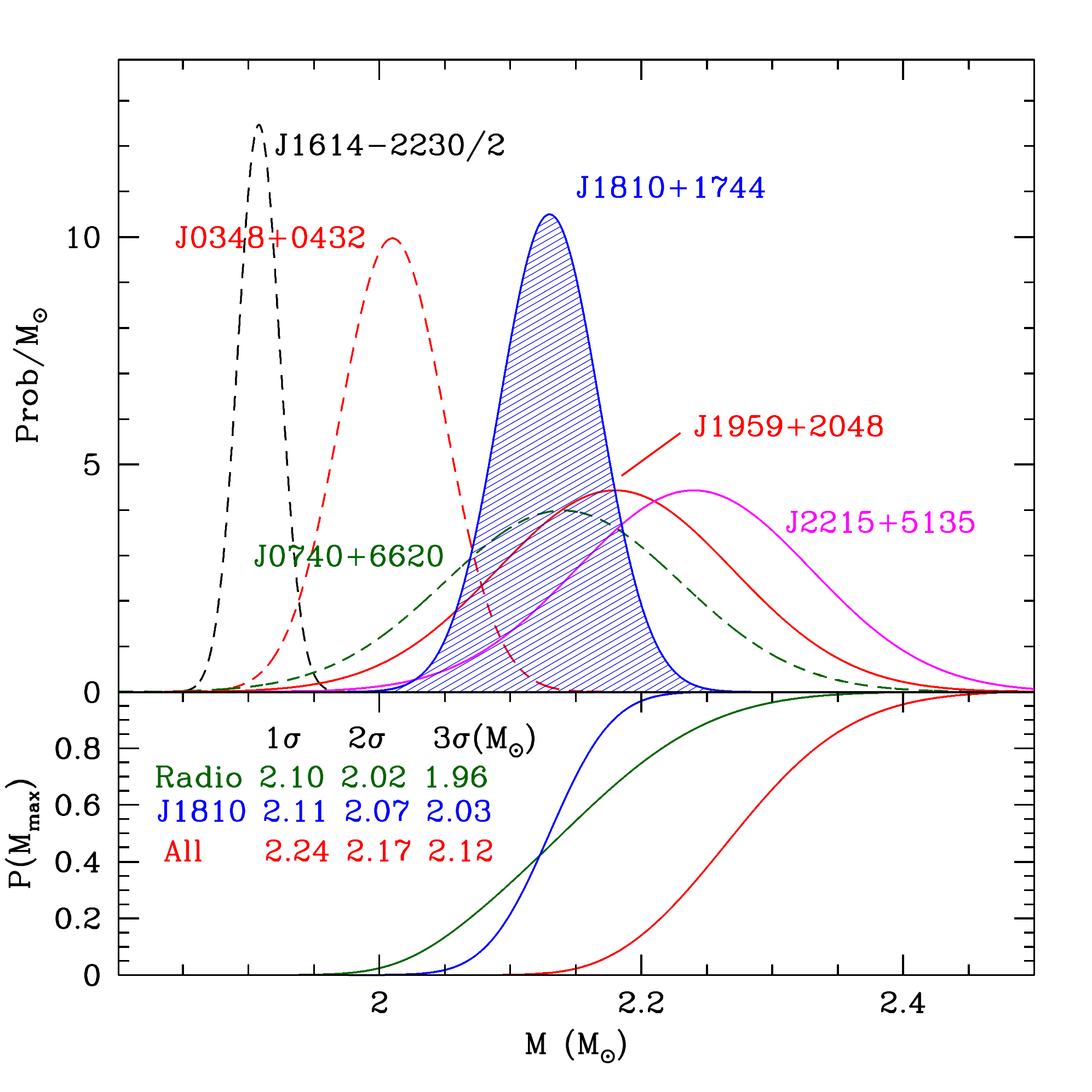}
\caption{Mass estimates for heavy NSs. The dashed curves show three radio-selected pulsars with white dwarf (WD) companions, measured from pulse timing (supplemented by WD atmosphere modeling for J0348). J1614 (amplitude decreased by a factor of 2 for plot) no longer significantly adds to the $M_{\rm max}$ lower bound. The solid curves show three spider binary mass estimates, relying on companion spectrophotometry. The well-determined J1810 mass range is shaded. The bottom panel shows the cumulative probability distributions for $M<M_{\rm max}$, for the radio objects, J1810 alone, and all six pulsars. Lower bounds ($1\sigma$, $2\sigma$, and $3\sigma$) are listed for these distributions.}
\label{fig:masses}
\end{figure}

The existence of heavy ($\sim 2\,M_\odot$) NSs was assured by the Shapiro delay measurement of PSR J1614$-$2230 \citep{demorest2010two}, although subsequent observations have refined and lowered the mass to $1.908\pm0.016\, M_\odot$ \citep{arzoumanian2018nanograv}. Two other mass measurements have proved very influential in the literature: PSR J0348+0432 at $2.01\pm0.04 \, M_\odot$ \citep{antoniadis2013massive} and the recent Shapiro-delay measurement of PSR J0740+6620 at $2.14^{+0.10}_{-0.09}\, M_\odot$ \citep{cromartie2020relativistic}, with all uncertainties $1\sigma$. By making a good-precision measurement of $M_{\rm NS}$ for J1810, we improve the lower bounds on $M_{\rm max}$. Indeed, this is the first individual object for which a $3\sigma$ lower bound on the mass exceeds $2\,M_\odot$. We can supplement these masses with the measurements for two other spider binaries from \citetalias{Kandel_2020}: PSR J1959+2048 at $2.18\pm0.09\, M_\odot$ and PSR J2215+5135 at $2.24\pm0.09\, M_\odot$ (although that analysis did not include GD effects so the actual inclinations and masses could be somewhat lower). Figure \ref{fig:masses} shows the mass uncertainty ranges for these objects.

Several approaches can be used to estimate $M_{\rm max}$. One option is to model the full distribution of (binary) NS masses and see if an upper cutoff is required; \citet{alsing2018evidence}, for example, determine that $M_{\rm max}$ is within a $1\sigma$ range of 2.0--2.2\,$M_\odot$. Here we only attempt to determine a lower bound to $M_{\rm max}$, so with individual source probability density functions (PDFs) $P_i = (2\pi\sigma_i^2)^{-1/2} e^{-0.5[(m_i-m)/\sigma_i]^2}$, we can form the joint probability of getting the set of measurements $\{d_i\}=\{m_i,\sigma_i\}$ when $M<M_{\rm max}$ as $\Pi_i \int_0^{M_{\rm max}} P_i\,\mathrm{d}m$. This is also the Bayesian probability $P(M_{\rm max} | d_i)$ for a flat prior with a hard cutoff, $\Theta (m- M_{\rm max})$. These are shown in the lower panel of Figure \ref{fig:masses}.

As demonstrated in this paper, we have marginalized over all parameters in determining the spider mass-estimate uncertainties. We have also explored alternative models; in most cases these (worse-fitting) models require {\it higher} masses. Since the spider binary mass estimates rely on the heating model, we cannot claim that they are free of systematic effects. However, we have attempted to be conservative --- and as more spider pulsars are measured with high masses and increasing accuracy, we should not ignore their contribution to constraints on $M_{\rm max}$ and thus on the dense-matter equation of state. Taken at face value, the uncertainties in recent spider measurements are sufficiently small  to significantly improve the bounds. We can now say with high ($3\sigma$ one-sided lower bound) statistical confidence that $M_{\rm max} > 2.12\,M_\odot$, and that at $\sim 1\sigma$ significance $M_{\rm max}> 2.24\,M_\odot$ is preferred. Additional measurements of spider binaries, especially the extreme BWs, will tighten (and likely slightly raise) this lower bound. 

In summary, we have found a good photometric+spectral model for BW J1810. This relies on substantial GD on the heated (day side), especially near the companion nose. Also important are HS and limb-darkening effects, particularly near binary minimum. These effects all serve to increase the best-fit inclination $i$ and to lower the inferred NS mass. Other binaries should be checked for this GD effect, but few have the large fill factor and high $T$ that make GD so strong in J1810. However, even after modeling this effect, J1810's NS mass is large, and alternate (worse-fitting) models tend to be even heavier. With our excellent fit precision, J1810 provides for the first time a lower limit on an NS mass that is greater than $2\,M_\odot$ at $>3\sigma$ confidence. This seems robust to any residual systematics and should thus be important for discussions of the dense-matter equation of state.
\bigskip

We thank the anonymous referee for a careful reading of the text. We are grateful for the excellent assistance of the staffs of the observatories where data were taken. Some of the data presented herein were obtained at the W. M. Keck Observatory, which is operated as a scientific partnership among the California Institute of Technology, the University of California, and NASA; the observatory was made possible by the generous financial support of the W. M. Keck Foundation.   
D.K. and R.W.R. were supported in part by NASA grants 80NSSC17K0024 and 80NSSC17K0502. A.V.F.'s group is grateful for generous financial assistance from the Christopher R. Redlich Fund, the TABASGO Foundation, and the Miller Institute for Basic Research in Science (U.C. Berkeley; A.V.F. is a Miller Senior Fellow).

\bibliographystyle{aasjournal}
\bibliography{main}
\end{document}